\documentclass[aps,prb,twocolumn,amsmath,amssymb,amsfonts,longbibliography,
    superscriptaddress,nobibnotes]{revtex4-2} 
\usepackage[utf8]{inputenc}
\usepackage[T1]{fontenc} 
\usepackage{ebgaramond}
\usepackage{braket}
\usepackage{graphicx} 
\usepackage{times}
\usepackage{soul}
\usepackage{url}
\usepackage{mathptmx}
\usepackage{gensymb}
\usepackage[
hypertexnames=false,
colorlinks=true,
citecolor=blue,
linkcolor=blue,
urlcolor=blue]
{hyperref}

\usepackage{float}

\usepackage{xcolor}
\definecolor{mygreen}{rgb}{0.0, 0.6, 0.0}

\DeclareMathAlphabet\mathbfcal{OMS}{cmsy}{b}{n}

\graphicspath{{fig/}{./fig/}{.}}

\makeatletter
\renewcommand\@make@capt@title[2]{%
 \@ifx@empty\float@link{\@firstofone}{\expandafter\href\expandafter{\float@link}}%
  {\textbf{#1}}\@caption@fignum@sep#2\quad
}%
\makeatother

\begin{document}

\title{Magnetic properties of moir\'e quantum dot arrays}

\author{Weronika Pasek}
\email[e-mail: ]{weronika.pasek@doktorant.umk.pl}
\affiliation{Institute of Physics, Faculty of Physics, Astronomy and Informatics, Nicolaus Copernicus University, Grudziadzka 5, 87-100 Toru\'n, Poland}
\author{Michal Kupczynski}
\affiliation{Department of Theoretical Physics, Wroclaw University of Science and Technology, 50-370 Wroclaw, Poland}

\author{Pawel Potasz}
\affiliation{Institute of Physics, Faculty of Physics, Astronomy and Informatics, Nicolaus Copernicus University, Grudziadzka 5, 87-100 Toru\'n, Poland}

\date{\today}

\begin{abstract}
We investigate magnetic properties of quantum dot arrays of moir\'e triangular superlattices. Starting from a reciprocal space model, we use the projection technique to obtain maximally localized Wannier functions and determine generalized Hubbard model parameters. The many-body Hamiltonian is solved using the exact diagonalization method as a function of the number of electrons in differently shaped quantum dots arrays. Finite spin polarization is observed within a wide range of filling factors for small twist angles and sufficiently strong interactions in most of the studied structures. The prospect for a magnetization controlled by applying a displacement field is presented. In the vicinity of half-filling, signatures of Nagaoka ferromagnetism in moir\'e materials are seen, which we demonstrate by comparing results with the corresponding on-site Hubbard model.
\end{abstract}

\maketitle

\section{Introduction}
\label{sec.intro}
Superlattices were shown to be a promising new platform for quantum simulators \cite{Review1, Review2, FengchengHubbard}. In particular, when transition metal dichalcogenides (TMD) heterobilayer are stacked, a difference in lattice constant or a twist between the layers produces a moir\'e superlattice.
The carriers at the top of the valence band of one of the layers experience a periodic potential formed by the other layer and many-body physics is expected to be described by a triangular lattice Hubbard model \cite{FengchengHubbard}. 

Correlated insulating states at half-filling \cite{HubbardCornell,HubbardETH,Berkeley,HubbardColumbia} with antiferromagnetic Curie–Weiss behavior \cite{HubbardCornell} and discrete series of insulating states identified as the generalized Wigner crystal states for partial fillings \cite{Berkeley, CornellWigner, CaliforniaWigner, ContinuousWigner} have been observed. In particular, Wigner crystal states at the $\nu = 1/3$ and $\nu = 2/3$ and the stripe phase at the $\nu=1/2$ filling were detected using non-invasive Scanning Tunneling spectroscopy (STS) imaging \cite{STM_Wigner} and optical anisotropy measurement \cite{CornellWignerStripe}. 

Recently, the importance of nonlocal interaction terms due to the finite high of the modulation potential strength was established \cite{Nonlocal,FrustratedWignerMott}. The role of different interaction terms in determining the ground state properties of the systems  
depends on a moir\'e potential depth, a moir\'e lattice constant determined by a twist angle,  background dielectric screening, and a filling factor, and all of these parameters can be experimentally controlled \cite{HubbardCornell,CornellWigner,CornellWignerStripe,CaliforniaWigner,FrustratedWignerMott, ExperimentMIT1}.

%, with the potential for engineering a variety of exotic correlated phases. 
While infinite moir\'e systems have been intensively studied, the physics related to their finite-size fragments have been considered only in a very limited context \cite{YongxinAllanpaper}. A single moir\'e localization potential for small twist angles can be approximately described by a harmonic oscillator \cite{FengchengHubbard,Nonlocal}, similarly to confining potentials in semiconductor quantum dots \cite{SpringerQuantumDots}. Quantum dots, also known as artificial atoms, are nanoscopic objects that can be considered as building blocks for artificial molecules. Different shape and size nanostructures with tunable hopping amplitude and interaction strength allow one to probe fermionic many-body physics \cite{salfi2016quantum,hensgens2017quantum,dehollain2020nagaoka}. Recently a few semiconductor quantum dot array patterned using the
scanning tunneling microscope (STM)-based hydrogen lithography technique has been demonstrated and proposed as Fermi-Hubbard model simulator \cite{BryantQuantumDots}. The field of artificial lattices rapidly develops and the formation of artificial molecules and superlattices has been proposed and demonstrated in many semiconductor materials \cite{singha2011two,gomes2012designer,PhysRevLett.111.185307,PhysRevLett.113.196803,drost2017topological,slot2017experimental,forsythe2018band,park2009making,wang2018observation,PhysRevB.104.245430,PhysRevB.105.205105}. Within several advantages of solid state physics nanostructures over other platforms for quantum simulations such as ultracold atoms in optical lattices \cite{bloch2005ultracold,ColdAtoms,mazurenko2017cold,gross2017quantum} are easy access to transport measurements, and dynamic control of the chemical potential landscape and filling factors using gates. 

%In general, in few electron nanostructures, magnetic and electronic properties can be controlled more accurately than in systems with a macroscopic number of particles. 
% allows to design of single/few electron electronic devices in the future. 

 In this work, we study a new type of artificial molecules created from moir\'e triangular superlattices and we focus on analysis of their magnetic properties. Quantum dot arrays of various shapes are considered and filling-factor dependent properties are determined. While a triangular lattice at the half-filling is not expected to reveal finite spin polarization within the Hubbard model, we show that moir\'e many-body Hamiltonian lead to a more complex magnetic phase diagram with a twist angle and interaction strength dependence. We complement our studies by analyzing the exchange interaction of an effective spin model which is also valid for infinite systems. Close to half-filling we investigate a contribution to magnetization coming from itinerant electrons and discuss the potential observation of Nagaoka ferromagnetism. For fillings away from the half-filling, shape-dependent charge orders are identified, in particular, Wigner molecules for triangular shape structures. We focus on finite size fragments of WSe$_2/$WS$_2$ heterostructure but our conclusions are valid for other heterobilayer TMDs.  

\section{Model and Methods}
\label{sec.Model}
We start from a continuum model to obtain Bloch states of moir\'e bands of holes (see Appendix). As shown in Refs. \cite{FengchengHubbard,potaszMetaInsul} the topmost moiré miniband is separated from other bands by the energy gap justifying the restriction of the Hilbert space to only that band. We use the projection technique to obtain corresponding exponentially localized Wannier functions \cite{Cloizeaux1,Cloizeaux2,Vanderbilt}. Next, we calculate real-space Coulomb matrix elements for a generalized Hubbard model. We include on-site and all direct Coulomb interactions, and non-local interaction terms - exchange, and assisted hopping - because they can play a crucial role in determining many-body properties of the ground states \cite{Nonlocal}. We will consider finite-size fragments of moir\'e superlattices with a given number of lattice sites. Hamiltonian for these moir\'e quantum dots arrays is written as
 \begin{equation}
 \begin{split}
    H  = & -\sum_{n=1}^3  t_n \sum_{<i,j>_n,\sigma} a_{i,\sigma}^\dagger a_{j,\sigma} +  U_0\sum_{i}n_{i,\downarrow}n_{i,\uparrow} \\ &- X_1\sum_{<i,j>,\sigma} n_{i,\sigma} n_{j, \sigma}+ A_1 \sum_{<i,j>,\sigma} (n_{i,-\sigma}+n_{j,-\sigma})a_{i,\sigma}^\dagger a_{j,\sigma}\\
    &+ \sum_{n=1}^3 U_n\sum_{<i,j>_n,\sigma,\sigma'} n_{i,\sigma} n_{j, \sigma'} ,
 \end{split}
 \label{sec.Ham}
\end{equation}
where $a_{i,\sigma}^\dagger$ ($a_{i,\sigma}$) is a fermionic operator, which creates (annihilates) electron with spin $\sigma$ on the lattice site $i$, $<i,j>_n$ depicts a pair of $n$-th nearest neighbor sites, and $i>j$ to avoid double counting in summations. Model parameters are $t_n$ hopping, $U_0$ on-site interaction, $U_n$ direct interaction, nearest-neighbor $X_1$ direct exchange, and $A_1$ assisted hopping. It is expected that the effective dielectric constant $\epsilon$ determining the interaction strength (see the Appendix) lies within a range $ 10 < \epsilon < 20$. Thus, we mainly investigate the properties of finite-size systems for two values determining these limits. This estimation is based on the dielectric constant of the electrostatic environment when hexagonal boron nitride is used as the substrate, $  \epsilon \approx 6$ \cite{stier2016probing} and additional screening by conducting gates and virtual transitions between the considered energy band and energetically remote moir\'e energy bands \cite{Nonlocal}.

Calculated Hamiltonian parameters for three different twist angles are shown in Table \ref{table:1} in the Appendix. Hamiltonian given by Eq. (\ref{sec.Ham}) is diagonalized in a basis of configurations corresponding to all possible distributions of particles on lattice sites. The ground state is characterized by its total spin $S$. 

\begin{figure}
    \centering
    \includegraphics[width=\columnwidth]{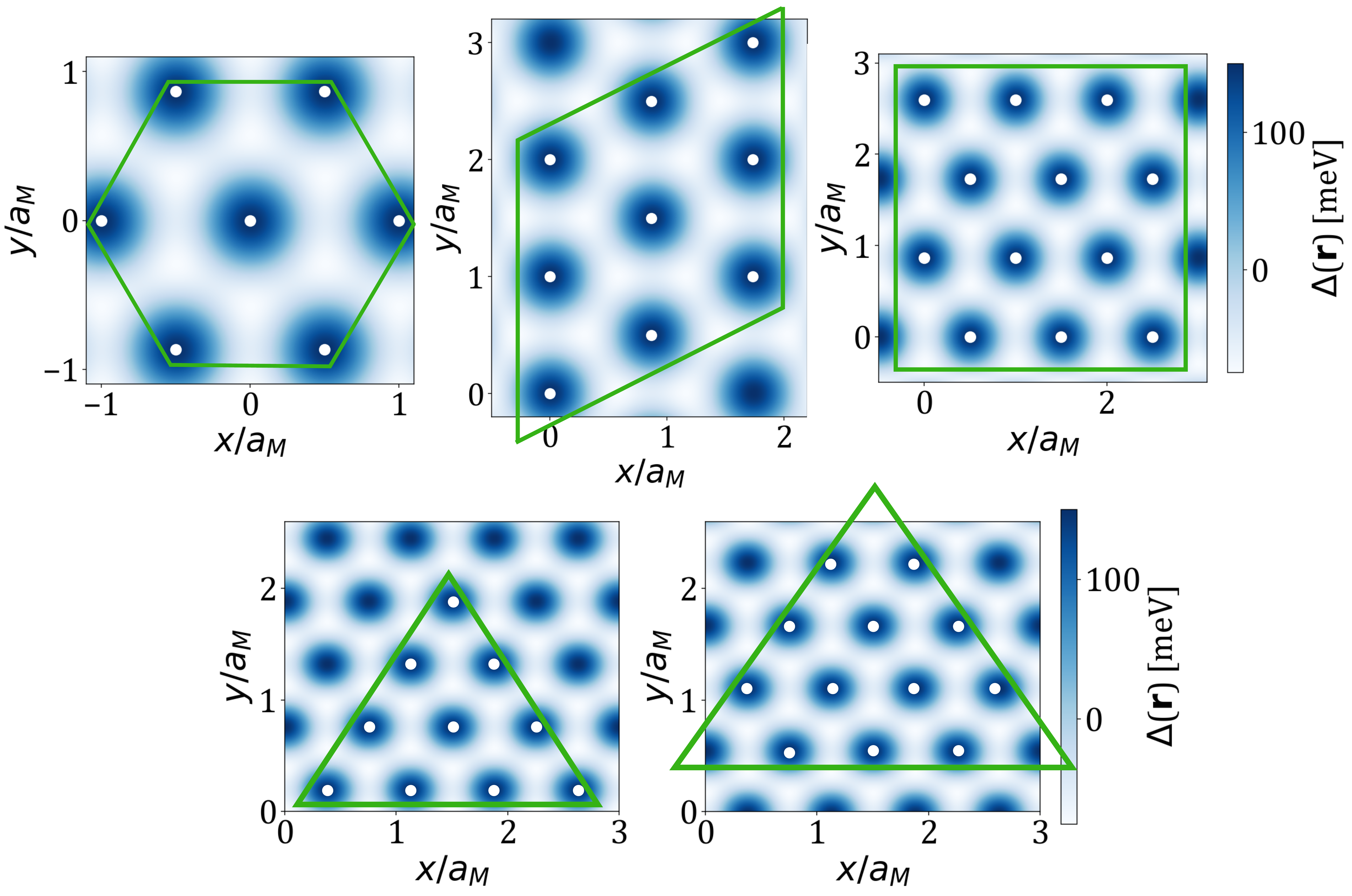}
    \label{fig:7-sites}
    \caption{
    The effective moir\'e potential forms a triangular lattice with lattice constant $a_M$, which depends on the value of the twist angle. Five analyzed structures consist of, in an upper row: $N=7$ (left), $N=9$ (middle), $N=12$ (right), and in a lower row: $N=10$ (left) and $N=12$ (right) moir\'e quantum dots. White dots indicate sites forming a given structure.}
    \label{fig:Fig1}
\end{figure}

\section{Magnetic phase diagram for $N=9$ sites structure}
\label{sec.PhaseDiag}

\begin{figure}[ht]
\begin{center}
\includegraphics[width=1.0\columnwidth]{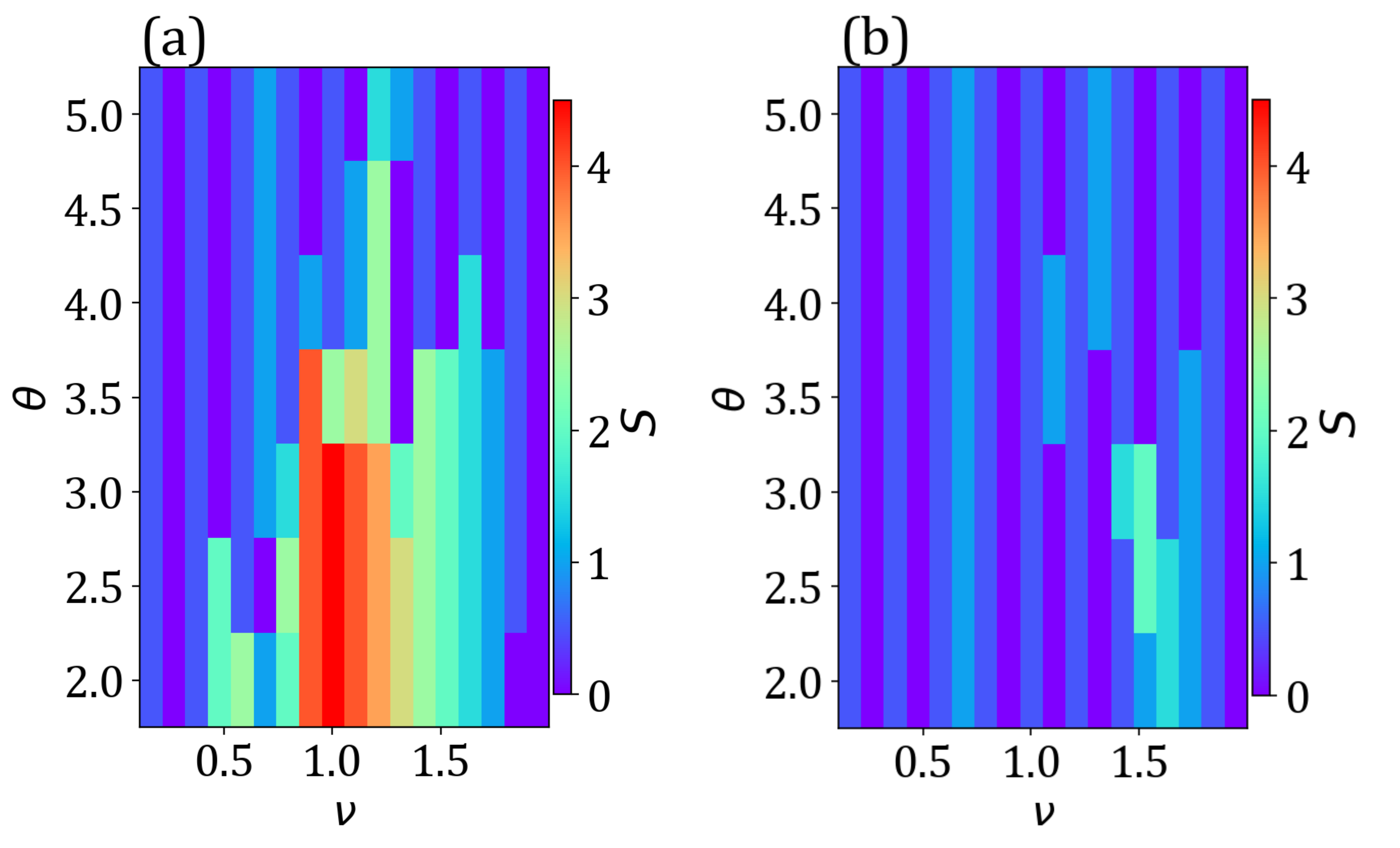}
\caption{The magnetic phase diagram of $N=9$ moir\'e quantum dot array. Total spin of the ground state as a function of the twist angle $\theta$ and filling factor $\nu$ for fixed values of the dielectric constant (a) $\epsilon=10$ and (b) $\epsilon=20$. }  
\label{Fig:Fig2}
\end{center}
\end{figure}

We study the magnetic properties of quantum dot arrays of different shapes shown in Fig. \ref{fig:Fig1}. Our representative example is the structure with $N=9$ quantum dots shown in Fig. \ref{fig:Fig1} in the middle of the top row. The total spin $S$ of the ground state is determined in a wide range of the twist angles, $2.0 \leq \theta \leq 5.0$, and filling factors $0 \leq \nu \leq 2$, where $\nu=N_{p}/N$ with $N_{p}$ the number of particles. A color map of the total spin is shown in Fig. \ref{Fig:Fig2}(a) for the effective dielectric constant $\epsilon=10$ and in Fig. \ref{Fig:Fig2}(b) for $\epsilon=20$.
For $\epsilon=10$ maximal spin polarization appears for twist angles below $\theta=4.0$ within a wide range of filling factors. The magnetic phase diagram is asymmetric with respect to half-filling, (seen clearly in the vicinity of smaller filling factors) which reveals
the fact that a triangular lattice is not bipartite (nonzero value of hopping $t_2$ and $t_3$ is also essential here). A mechanism of vanishing spin polarization for larger twist angles is related to an increase in single-particle energy levels separations, which are too large compared to effective exchange interaction.  
The phase diagram for weaker interaction strength, $\epsilon=20$, shown in Fig. \ref{Fig:Fig2}(b), in general does not reveal spin polarization. The energy level separations are too large here compared to Coulomb energy scales and particles doubly occupy the lowest energy states. 

\subsection{Filling factor $\nu=1$}
\label{sec.Nu1}
A half-filled triangular lattice is expected to have a three sublattice antiferromagnetic order within the on-site Hubbard model and a limit of strong interaction \cite{NeelTrianElser, NeelTrianWhite, PhysRevLett.100.076402,PhysRevLett.103.036401,RevModPhys.90.025003,PhysRevLett.127.087201}.
A finite magnetization noticeable in Fig. \ref{Fig:Fig2}(a) for $\epsilon=10$ at the half-filling is related to the nonzero value of direct exchange interaction due to overlap of neighboring Wannier functions, as a consequence of finite height of moir\'e localization potentials \cite{Nonlocal}. This direct exchange interaction has to be sufficiently large to overcome an energetic cost of the occupation of higher energy states, but additionally, competes with an antiferromagnetic superexchange interaction. The physics at half-filling can be explained using an effective spin model for sufficiently strong interaction, as we will discuss below.

\subsubsection{Exchange $J$ of an effective spin model}
\label{section-spin-model}

\begin{figure}
    \centering
    \includegraphics[width=\columnwidth]{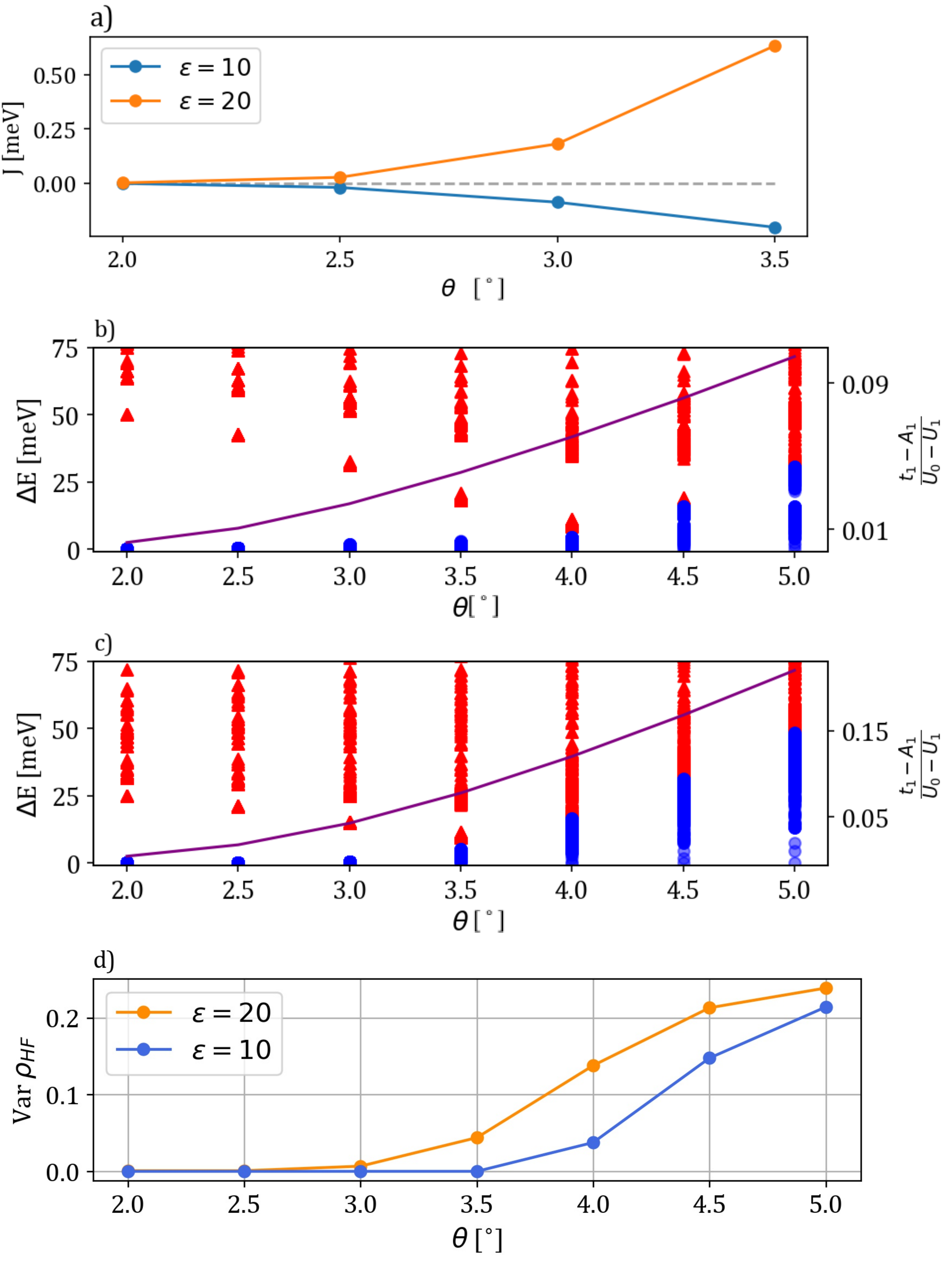}
    \caption{(a) The effective exchange interaction $J$ (Eq. \ref{Jterm}) for $\epsilon=10$ and $\epsilon=20$ as a function of the twist angle. (b,c) The energy spectrum of many-body Hamiltonian for half-filling, $\nu=1$, measured from the ground state $\Delta E= E_i - E_0$ for $N=9$ moir\'e quantum dot array as a function of a twist angle for (b) $\epsilon=10$ and (c) $\epsilon=20$. Blue color marks the first $2^9=512$ states of the lower Hubbard band. A finite energy gap between lower and upper Hubbard bands approximately indicates regimes where the description of low energy physics using an effective spin model is justified. A continuous line corresponds to expression $\frac{t_1 + A}{U_0-U_1}$ with a scale on the right. (d) A variation from a single occupation of each quantum dot as a function of a twist angle.}   
    \label{Fig:Fig3}
\end{figure}

The spectrum of the Hubbard model in a limit of strong interaction separates into two bands, lower and upper Hubbard bands. The lower Hubbard band consists of $2^N$ singly occupied quantum dots, while the upper Hubbard band has some doubly occupied quantum dots. Because of that separation between the two bands is proportional to onsite interaction $U$. Projecting Hamiltonian onto the lower Hubbard band gives
effective spin model, the Heisenberg model in this case, with effective exchange interaction $J$ derived for a generalized Hubbard model in Ref. \cite{Nonlocal}: 
 
\begin{equation}
\begin{split}
&J = -2X_1 + \frac{4\tilde{t_1}^2}{U_0-U_1} + \frac{8\tilde{t_1}^4}{(U_0-U_1)^3} \left( \frac{U_0 - U_1}{2U_0-3U_1+U_2} \right. \\ 
&+ \left. \frac{4(U_0-U_1)}{2U_0-U_1-U_2} + \frac{3(U_0-U_1)}{U_0-U_2} + \frac{2(U_0-U_1)}{U_0-U_3} - 11  \right),
\end{split}\label{Jterm}
\end{equation}
where $\tilde{t_1} = t_1 - A_1$. In Fig. \ref{Fig:Fig3}(a) we show $J$ for twist angles below $\theta = 4.0$. A negative value of $J$ indicates a ferromagnetic state and a positive one indicates an antiferromagnetic order. A sign of $J$ agrees with magnetic phase diagrams at half-filling $\nu=1$ for moir\'e quantum dot arrays shown in Fig. \ref{Fig:Fig2}(a) for $\epsilon=10$. For $\epsilon=20$, no finite spin polarization is expected, $J$ is always positive, and this agrees with results shown in Fig. \ref{Fig:Fig2}(b). 

While according to the behavior of $J$, spin polarization should sustain for larger angles $\theta > 3.5$, the effective spin model is not valid any longer in that regime. This is shown in Fig. \ref{Fig:Fig3}(b) and \ref{Fig:Fig3}(c), where we analyze the energy gap between the upper and lower Hubbard bands. The lower Hubbard band consists of $2^N = 512$ spin states (N=9), indicated by blue circles in Fig. \ref{Fig:Fig3}(b,c). The gap to higher energetic states from the upper Hubbard band (red triangles) is visible for small twist angles. The gap closes when the twist angle increases, vanishing around $\theta=4.0$ for $\epsilon=10$, and $\theta=3.5$ for $\epsilon=20$. These are roughly twist angles, where a spin model approximation breaks down. We can relate these estimations to the analysis of $\frac{t}{U}$ ratio from Ref \cite{effSpinModel} with a spin model applicability estimated for $\frac{t}{U}<0.15$ for the Hubbard model.  In our case, the energy gap between upper and lower Hubbard bands vanishes already for $\frac{t_1 + A}{U_0-U_1} \simeq 0.07$ for $\epsilon = 10$,  Fig. \ref{Fig:Fig3}(b), and for $\frac{t_1 + A}{U_0-U_1} \simeq 0.05$ for $\epsilon = 20$, Fig. \ref{Fig:Fig3}(c), while mentioned analysis does not take into account direct exchange $X$ interaction. The final estimation of spin model approximation can be done by looking at quantum dot occupations. In the spin model, each site is occupied by exactly one particle. A variation  from a single occupation is defined as
\begin{equation}
    \text{Var} \rho = \frac{1}{N} \sum_i^N |\rho^{E}_i - 1|,
\end{equation}
 where $\rho^{E}_i$ is the electron density at the $i$-th lattice site, and $\text{Var} \rho$ should be close to zero for spin model regime. $\text{Var} \rho$ as a function of a twist angle is shown in Fig. \ref{Fig:Fig3}(d) and conclusions agree with that from the energy gap between upper and lower Hubbard band estimations, assuming a threshold at $\text{Var} \rho = 0.01$ for a critical variation of single occupation.
While the results shown in Fig. \ref{Fig:Fig3}(b)-(d) regard the structure with $N=9$ quantum dots, the drawn conclusion is expected to be valid for arbitrary systems, including periodic structures.

\subsubsection{Charge and spin orders}
\begin{figure}[ht]
\begin{center}
\includegraphics[width=0.95\columnwidth]{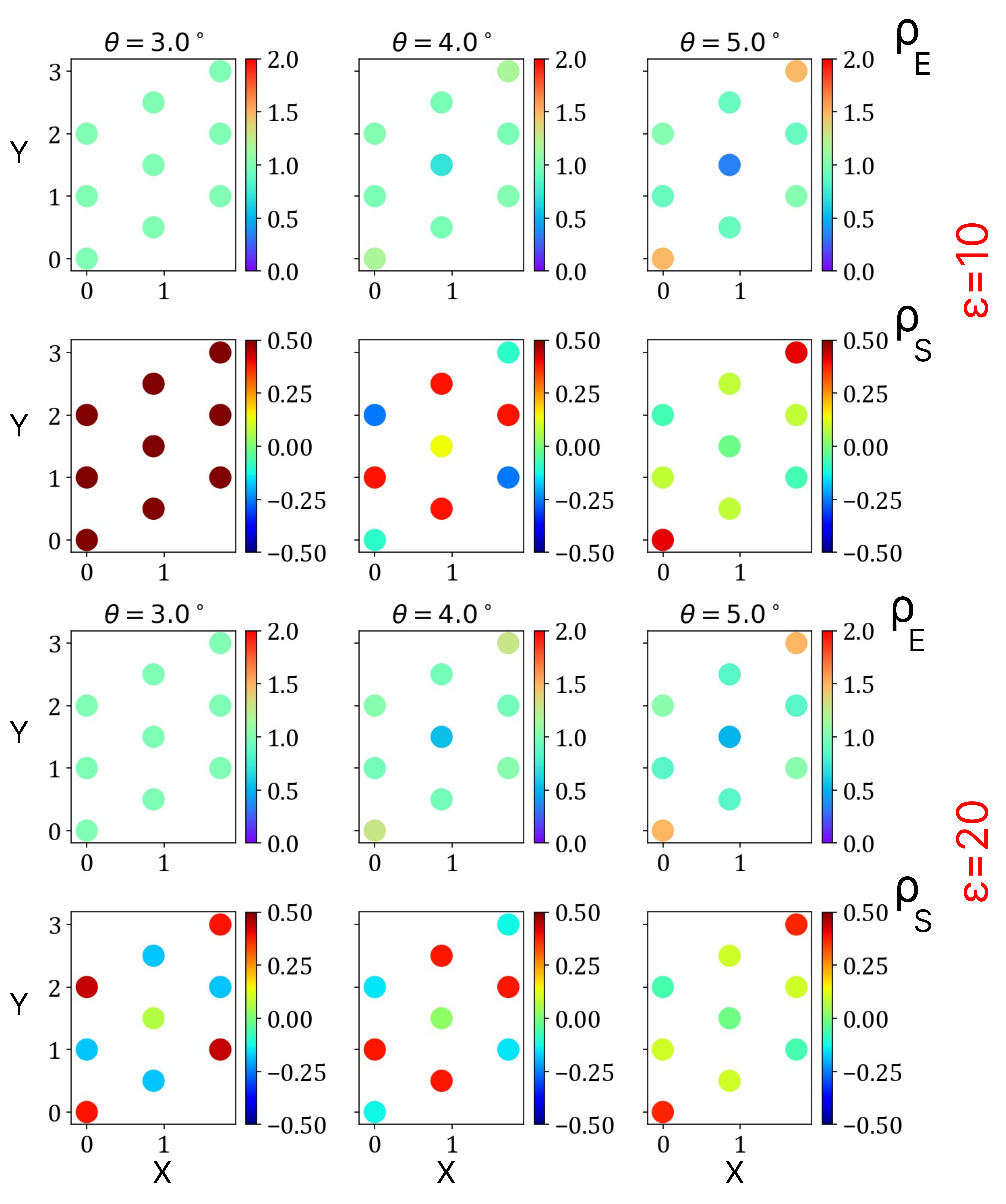}
\caption{The ground state charge (the first and the third row) and spin (the second and the fourth row) densities for the half-filling, $\nu=1$, for $N=9$ moir\'e quantum dot array. The results for twist angle $\theta = 3.0$ (left column), $\theta = 4.0$ (middle column), $\theta = 5.0$  (right column) and dielectric constant  $\epsilon=10$ (top two rows) and $\epsilon=20$ (bottom two rows).}
\label{Fig:FigDens}
\end{center}
\end{figure}

We investigate the charge and spin densities at the half-filling for three different twist angles and both values of the dielectric constant, shown in Fig. \ref{Fig:FigDens}. The top two rows show the charge and spin densities, distributed over quantum dots in the case of strong interactions, $\epsilon=10$, and the bottom two rows to a weaker interaction regime, $\epsilon=20$. For $\theta=3.0$ and $\epsilon=10$, spin and charge densities are uniform, as expected in a regime of the spin model and the fully spin-polarized ground state with $S=S_{max}$. With a larger twist angle, for both dielectric constants, the occupation of a central quantum dot gets smaller compared to eight quantum dots around it, which is related to the presence of repulsive direct Coulomb interaction in Hamiltonian given by Eq. \ref{sec.Ham}. The energy cost needed to populate the central dot is the highest because of interactions with electron in the six nearest neighbor dots. One can expect that when charging the system, electrons first occupy external quantum dots and at the end the central dot. For $\theta=5.0$, the central dot is almost empty and the highest electron density is at the two quantum dots that are farthest away. The charge distribution in this case is similar for both values of the dielectric constant.

In the case of weaker interactions $\epsilon=20$, the spin density for $\theta=3.0$ with the ground states total spin $S=S_{min}$ is not uniform. Six quantum dots around the central one have positive spin density while the central dot and the two farthest away dots have negative spin densities. This spin order changes when a twist angle is increased to $\theta=4.0$. Now, spin densities resemble a stripe phase with uniform spin densities along a shorter axis of the quantum dot array with stripes of positive and negative spin densities. In general, spin densities reveal the two-fold rotational symmetry of the quantum dot array. Three sublattices' antiferromagnetic phase expected for an infinite triangular lattice can not be seen here because this type of the quantum dot array has too strong finite-size effects. 

\subsubsection{Magnetization controlled by a displacement field}

\begin{figure}[ht]
\begin{center}
\includegraphics[width=\columnwidth]{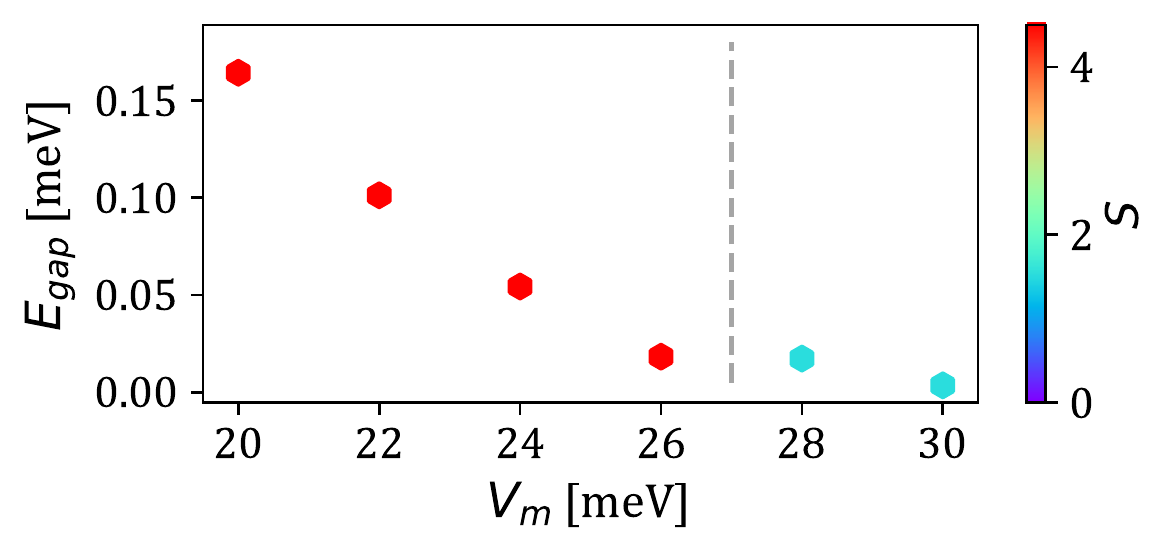}
\caption{The dependence of the energy gap between the ground state and the first excited state and magnetization on the depth of the effective moir\'e potential $V_{\rm m}$, which can be tuned by an external electric field perpendicular to the structure. Results obtained for the half-filling $\nu=1$, and parameters $\theta=3.0$, $\epsilon=10$ for $N=9$ moir\'e quantum dot array. A dashed line indicated a transition between the ground state total spin.}   
\label{Fig:Fig6}
\end{center}
\end{figure}

A displacement field has been used to control the moir\'e potential depth $V_{\rm m}$ and induce a metal-insulator transition \cite{ExperimentMIT1,ExperimentMIT2}.
 In Fig. \ref{Fig:Fig6} we show that it can be used to control the magnetic properties of quantum dot arrays at the half-filling. The ground state is a maximally spin-polarized state for parameters $\nu=1$, $\theta=3.0$, $\epsilon=10$, and $V_{\rm m}=25$ meV without a displacement field. Applying a displacement field can change the depth of a moir\'e localization potential $V_{\rm m}$ and effectively change the bandwidth. The energy gap between the ground state with maximal total spin increases its energy when $V_{\rm m}$ is increased, while the excited states with lower total spin decrease. After reaching a critical value, the total spin is lowered. The transition between the ground state total spin is indicated in Fig. \ref{Fig:Fig6} by a dashed line.  

\subsection{Vicinity of half-filling, $\nu=1,1^\pm$}
\begin{figure}[ht]
\begin{center}
\includegraphics[width=\columnwidth]{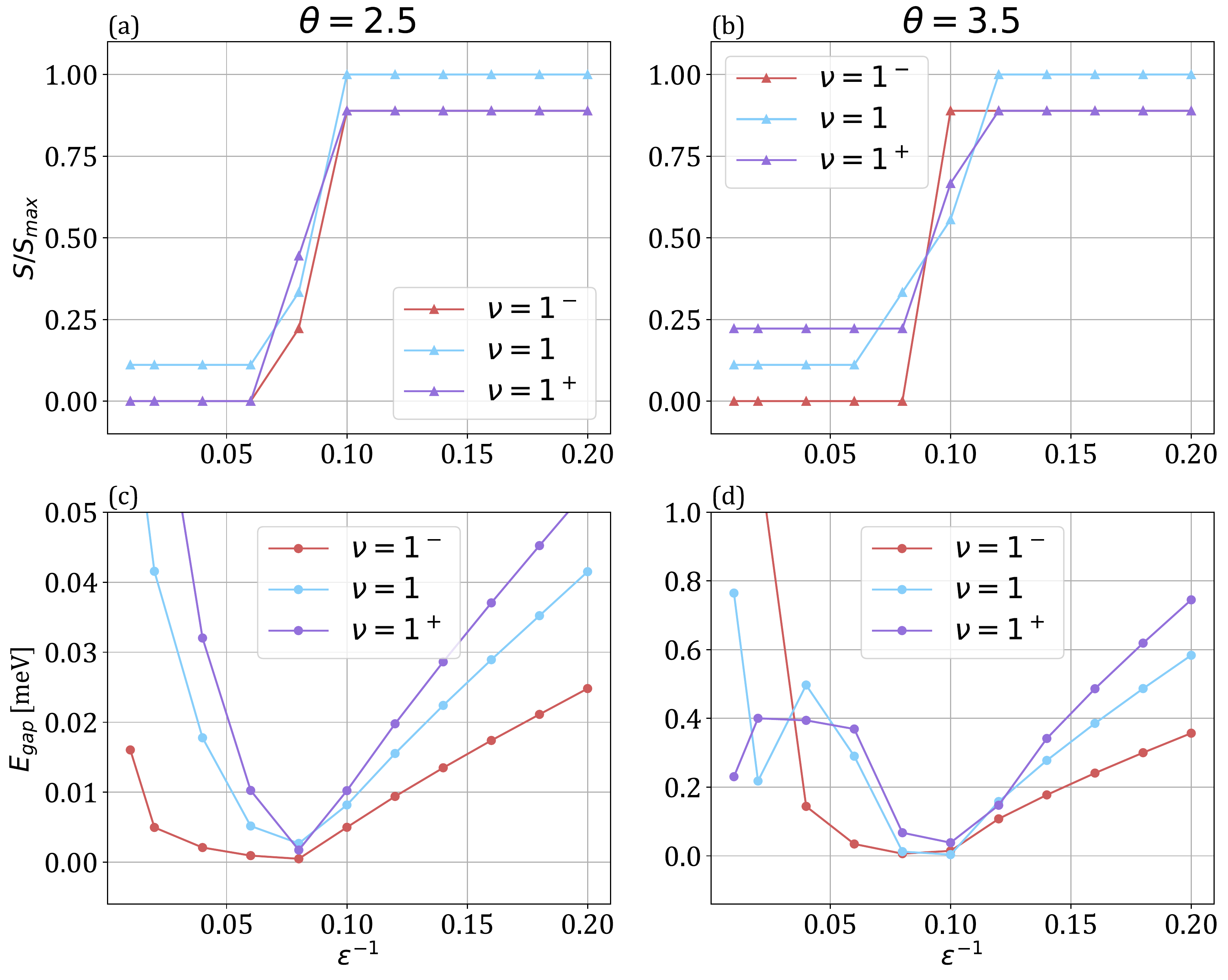}
\caption{The dependence of the total spin and energy gap on a dielectric constant near half-filling for two values of the twist angle, (a,c) $\theta=2.5$ and (b,d) $\theta=3.5$ for $N=9$ moir\'e quantum dot array. A transition to a spin-polarized phase is seen around $\epsilon\approx 10$ for both angles.}   
\label{Fig:Fig7}
\end{center}
\end{figure}

As we indicated previously, the ground state is maximally spin-polarized in the vicinity of the half-filling $\nu=1,1^\pm$ for $\epsilon=10$ and twist angles $\theta < 4.0$ (Fig. \ref{Fig:Fig2}), where $1^\pm$ labels the half-filling with extra $\pm$ one particle. In Fig. \ref{Fig:Fig7} we analyze the total spin and the energy gap between the ground state and the first excited state as a function of the interaction strength $\epsilon$ for fillings $\nu=1,1^\pm$ and for the twist angles $\theta=2.5$ and $\theta=3.5$. 
A transition from the ground state with total spin $S=S_{\rm min}$ to a state with total spin $S=S_{\rm max}$ occurs near $\epsilon^{-1}=0.08$ for both twist angle and for all three fillings considered here. Above this critical value of interaction strength, effective exchange interaction dominates and the polarization of spins is favored. Panels (c) and (d) of Fig. \ref{Fig:Fig7} show the evolution of corresponding energy gaps between the maximally spin-polarized ground state and the first excited states with a lower total spin. After the transition to a magnetic phase, the gap is the largest for filling $\nu=1^+$. We relate it to the kinetic mechanism of spin polarization within the Hubbard model proposed by Nagaoka \cite{nagaoka1966ferromagnetism}, which we describe below. 

\subsection{Hubbard model and Nagaoka ferromagnetism for filling factor $\nu=1^+$}
\begin{figure}[h]
\begin{center}
\includegraphics[width=\columnwidth]{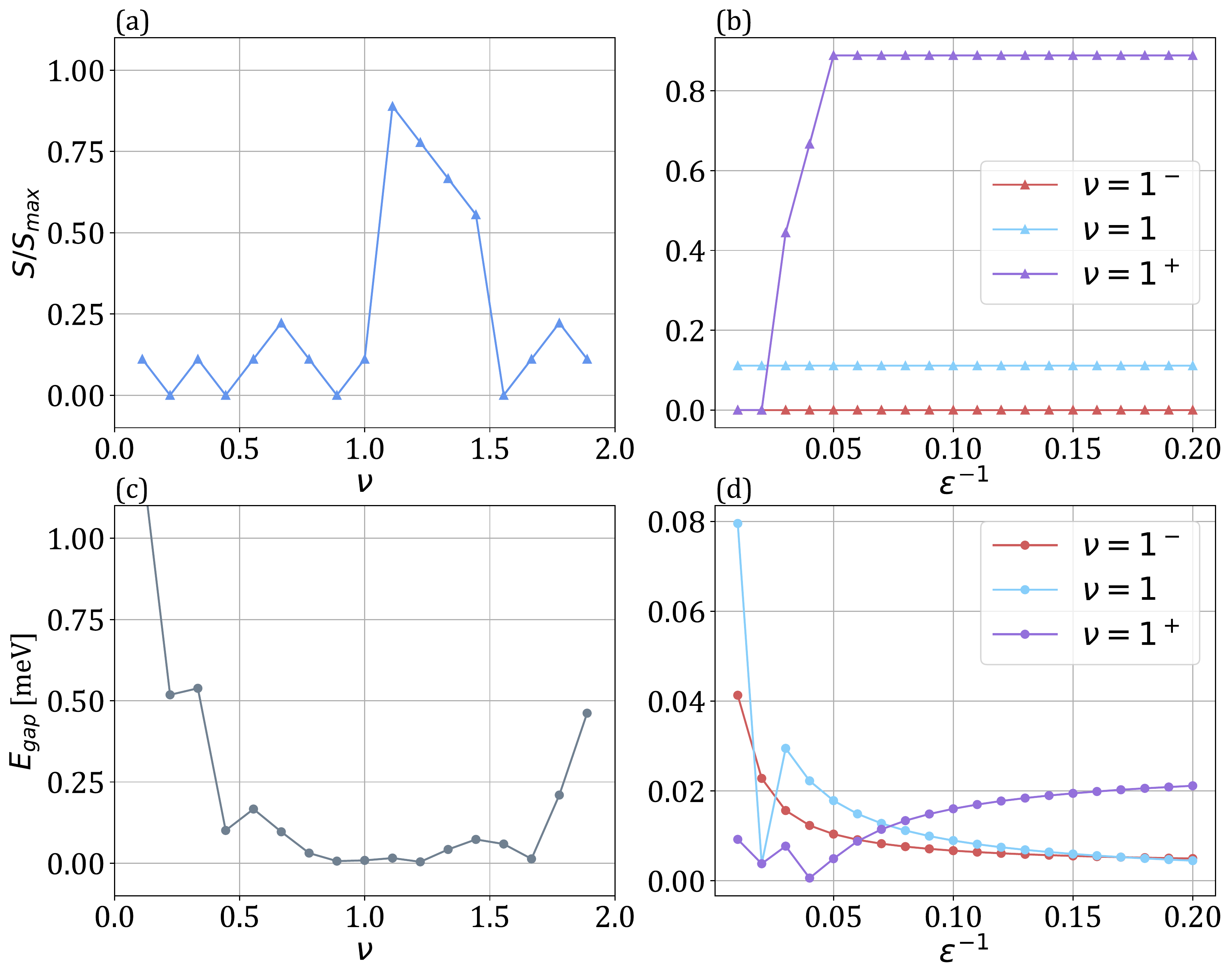}
\caption{Results obtained for the Hubbard model, Eq. (\ref{eq:hubbard}) for $N=9$ structure. (a) Total spin of the ground state and (c) the energy gap as functions of the filling factor for $\epsilon=10$ and $\theta=2.5$. (b) Total spin and (d) the energy gap for fillings $\nu=1^-, 1, 1^+$ and $\theta=2.5$ as a function of interaction strength.}
\label{Fig:Fig8}
\end{center}
\end{figure}

 We consider the on-site Hubbard model \cite{HubbardUterms} given by the Hamiltonian
\begin{equation} \label{eq:hubbard}
    H = - t_1 \sum_{<i,j>,\sigma} a_{i,\sigma}^\dagger a_{j,\sigma} + U_0\sum_{i}n_{i,\downarrow}n_{i,\uparrow}
\end{equation}
with parameters $U_0$ and $t_1$ taken from Table \ref{table:1} for twist angle $\theta=2.5$ and all other terms are neglected. This approximation can be justified by noting that nearest neighbor hopping $t_1$ and on-site interaction $U_0$ are significantly larger than other terms for this particular twist angle.
Fig. \ref{Fig:Fig8} shows the total spin of the ground state in panel (a) and the energy gap between the ground state and the first excited state in panel (c) as a function of the filling factor and for interaction strength $\epsilon=10$. There is a range of fillings $1 < \nu < 1.5$ with finite spin polarization. Values of energy gaps suggest stronger stability of magnetic phases closer to $\nu = 1.5$.  

In particular cases, when one electron or one hole is added to the half-filled system, $\nu=1$, a transition to a maximally spin-polarized state is expected within the Hubbard model in a limit of infinite $U_0$ interaction due to Nagaoka ferromagnetism \cite{Thoulessnagaokaferro,nagaoka1966ferromagnetism}. Whether on which side of the half-filling it occurs, for $\nu=1^-$ or $\nu=1^+$, depends on a lattice type. For a triangular lattice, Nagaoka ferromagnetism is expected for the Hubbard model for $\nu=1^+$  because a ferromagnetic state was shown to be unstable for $\nu=1^-$ concerning a Gutzwiller single spin flip \cite{hanisch1995ferromagnetism,hanisch1997lattice}. Indeed, we see it in Hubbard model results, which is the reason for the increased energy gap in moir\'e quantum dot arrays for this filling shown in Fig. \ref{Fig:Fig7}. We analyze the vicinity of half-filling,  $\nu=1,1^\pm$, when interaction strength is varied, showing in Fig. \ref{Fig:Fig8}(b) the total spin and in Fig. \ref{Fig:Fig8}(d) the energy gap between the ground state and the first excited state. A transition to the maximal spin-polarized ground state occurs only for $\nu=1^+$ for $\epsilon^{-1}\approx 0.05$. The energy gap increases with the increase of interactions strength but saturates at a constant value already around $\epsilon^{-1}\approx 0.15$. The kinetic mechanism responsible for finite spin polarization for $\nu=1^+$ within the Hubbard model contributes to spin polarization at this filling within the generalized Hubbard model, making it more stable compared to other fillings.

%\begin{figure}[ht]
%\begin{center}
%\includegraphics[width=0.95\columnwidth]{PBCvsOBC.png}
%\caption{The comparison of the total spin (a, c) and the energy gap (b,d) of the 9-site structure with periodic boundary conditions (PBC) and open boundary conditions (OPC). Results have been obtained for two values of the dielectric constant $\epsilon=10,20$ and fixed $\theta=3.0$. \WP{Plot fixed}}   
%\label{Fig:Fig9}
%\end{center}
%\end{figure}

\section{Shape dependent magnetization in moir\'e quantum dot arrays }
\label{sec.Differnet}

\begin{figure}[h]
\begin{center}
\includegraphics  [width=\columnwidth]{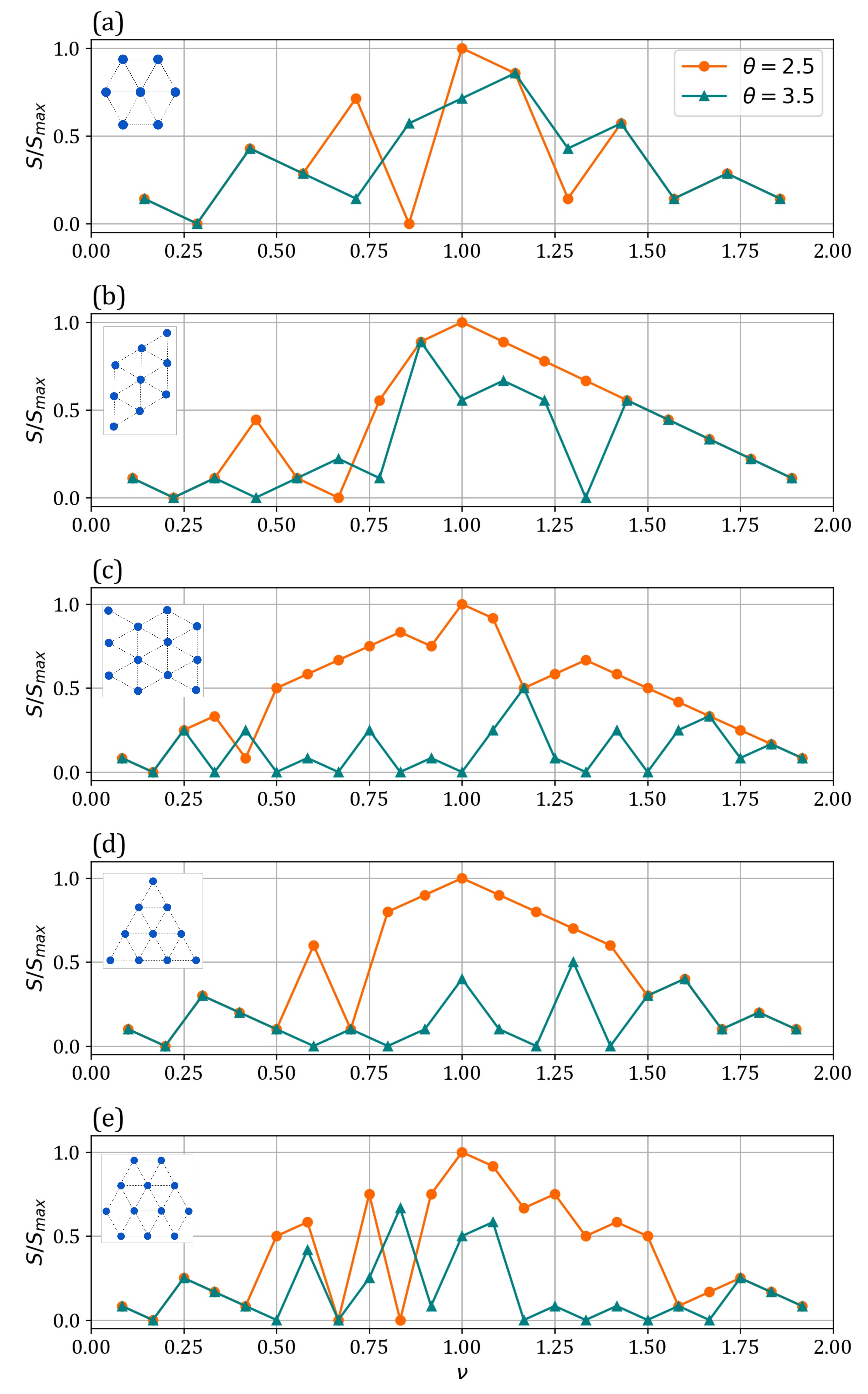}
\caption{The comparison of magnetic phase diagrams for moir\'e quantum dot arrays with (a) $N=7$, (b) $N=9$, (d) $N=10$ and (c,e) $N=12$. Insets in the upper right corners show the structure. The value of the dielectric constant is fixed to $\epsilon=10$.} 
\label{Fig:Fig10}
\end{center}
\end{figure}

In this section, we extend the previous analysis of magnetic properties to structures with various shapes, consisting of $N=7, 9, 10, 12$ moir\'e quantum dots, shown in Fig. \ref{fig:Fig1}. The total spin of the ground state as a function of the filling factor $\nu$ for $\epsilon=10$ and two twist angles, $\theta=2.5$ and $\theta=3.5$, is presented in Fig. \ref{Fig:Fig10}. We measure it in relation to its maximal value at the half-filling, $S_{\rm max} = \frac{N}{2}$. In general, a tendency for spin polarization is stronger for a smaller twist angle. For this twist angle, all structures have a maximal total spin at half-filling $\nu=1$, and for the filling with one extra electron added, $\nu=1^+$, where the Nagaoka mechanism of spin polarization plays a role. Finite spin polarization for half-filling, small twist angles, and sufficiently strong interaction seems to be independent of the shape and size of moir\'e quantum dots and agrees with expectations at the thermodynamic limit (from the sign of $J$, see Fig. \ref{Fig:Fig3}(a)). Additionally, in all structures, at least partial spin polarization occurs for filling factors in a range $1<\nu <1.5$, while below $\nu=1$ it is a structure-dependent, for example for $N=12$ triangular moir\'e quantum dot array spin polarization oscillates between a maximal and minimal value, Fig. \ref{Fig:Fig10}(e). When the twist angle is increased to $\theta=3.5$, a tendency for spin polarization decreases, appearing occasionally for some particular fillings. We now focus on symmetric (triangular) moir\'e quantum dots, where shape effects are stronger, revealing a geometry-related charge order.  
 
\subsection{Triangular Wigner molecules}

The shape of moir\'e quantum dot arrays determines their charge distribution when the filling factor is far from half-filling. Due to direct Coulomb interaction present in Hamiltonian given by Eq. \ref{sec.Ham}, the population of quantum dots with the largest separation is energetically favorable for such fillings. This leads to the observation of accumulation of charges in the corners of triangular moir\'e quantum dot, formation of Wigner molecules. This situation is similar to previously studied triangular graphene quantum dots \cite{potaszTGQD}.

\begin{figure}[h]
\begin{center}
\includegraphics[width=\columnwidth]{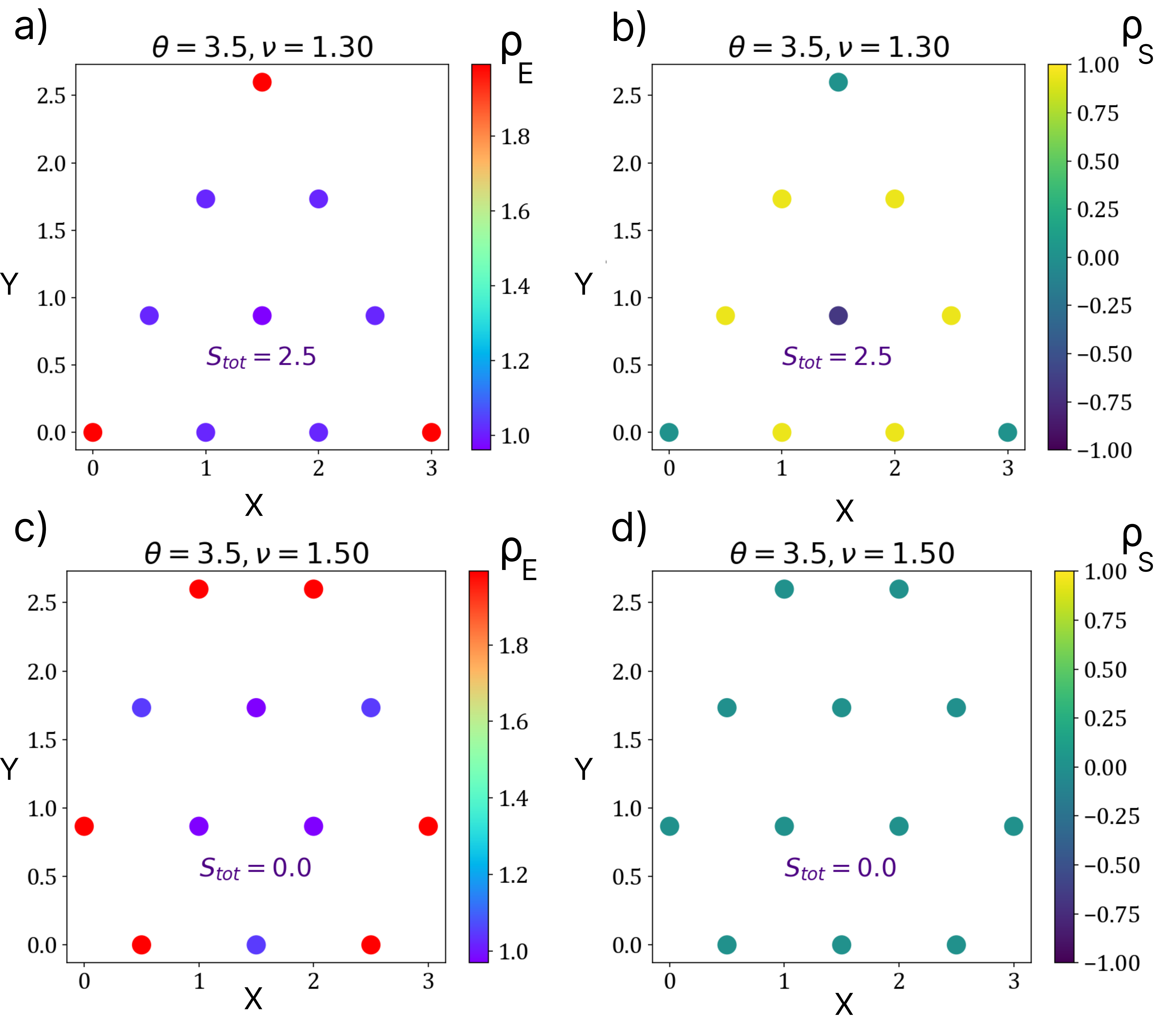}
\caption{Charge (a, c), and spin (b, d) densities of the ground state of triangular moir\'e quantum dot arrays with $N=10$ for $\nu = 1.3$ ($N_{\rm p}=13$), (a) and (b), and $N=12$ for $\nu = 1.5$ ($N_{\rm p}=18$), (c) and (d). The twist angle and the dielectric constant are fixed to $\theta=3.5$ and $\epsilon=10$.}
\label{Fig:Fig9}
\end{center}
\end{figure}

For the N=10 structure, shown on the left in a lower row in Fig. \ref{fig:Fig1}, Wigner molecules characterized by maximized charge density at three corners,  occur for the system with $N_{\rm p}=3,7,13,17$ particles. Two middle particle numbers correspond to $\nu=0.7$ and $\nu=1.3$, and equivalently, to removal/addition of $N_{\rm p}=3$ particles from/to the charge neutral system. In Fig. \ref{Fig:Fig9}(a) and Fig. \ref{Fig:Fig9}(b) we show the charge and spin densities of a representative example for $\nu=1.3$. Three corners are doubly occupied, and the rest of the moir\'e quantum dots are singly populated. The total spin of the ground state $S=2.5$ with a spin-down particle in the center and six spin-up particles around it. For the second triangular shape structure with $N=12$, shown on the right in a lower row in Fig. \ref{fig:Fig1}, we observe similar behavior after adding six particles to the half-filling, and symmetrically after adding six holes, $N_{\rm p}=18$ and $N_{\rm p}=6$, respectively, because here three corners are formed from two quantum dots. Electronic and spin densities of the former case are shown in Fig. \ref{Fig:Fig9}(c) and Fig. \ref{Fig:Fig9}(d). Total spin has minimal value and spin density is uniformly distributed. 

We notice here that Wigner molecules are observed regardless of the twist angle in both triangular structures, while their magnetic properties are twist angle dependent. For example, for $N=10$ structure, the ground state total spin for $\theta=2.5$ and $\nu=1.3$ is $S=3.5$ with all singly occupied quantum dots fully spin-polarized, while for $\theta=3.5$ the ground state total spin is $S=2.5$, with the spin of the center quantum dot flipped, as shown in Fig. \ref{Fig:Fig9}(b). A similar situation occurs for $N=12$ and $\nu=1.5$, with a change of spin polarization from maximal for $\theta=2.5$ to minimal for $\theta=3.5$.

\section{Comparison with periodic systems}
\label{sec.Periodic}
At half-filling, moir\'e quantum dot arrays have similar magnetic properties to periodic systems. A metal-insulator transition and a transition between antiferromagnetic and ferromagnetic order in infinite moir\'e superlattices have been predicted by one of us \cite{potaszMetaInsul,Nonlocal}. In agreement with these results is an observation of finite spin polarization in all moir\'e quantum dot arrays for strong interaction ($\epsilon=10$) and small twist angle ($\theta=2.5$) and no spin polarization when an interaction strength decreases or a twist angle increases. In the latter case, studied systems are too small to identify the expected long-range 120$^o$-N\'eel antiferromagnetically ordered state. Farther from half-filling, long-range direct interaction plays a major role in determining the charge order. In infinite systems, experimental and theoretical analysis have indicated many correlated insulating states away from the half-filling for $\nu = 1/3, 1/2, 2/3, 3/4$, and recognized them as general Wigner crystal phases \cite{CornellWigner, CaliforniaWigner, ContinuousWigner,GenWigCrysPadhi, LiangFuchargeorder}. Corresponding magnetic orders for filling $\nu = 1/3$ and $\nu = 2/3$ have been experimentally determined with proved antiferromagnetic order in the later one and no conclusive order in the former due to too small energy scale \cite{FrustratedWignerMott}. The magnetic honeycomb pattern found at $\nu = 2/3$ is in agreement with recent theoretical studies by one of us \cite{potaszWigner} with an additionally anticipated transition between antiferromagnetic and ferromagnetic orders. While, the charge orders for these partial fillings observed in periodic systems can not be expected in moir\'e quantum dot arrays because of geometric restrictions, here instead Wigner molecules are observed, and a transition between finite and no spin polarization with an increase of the twist angle is quite common for many fillings and in most of the structures, see Fig. \ref{Fig:Fig10}.

\section{Conclusions}
We investigated the magnetic properties of various moir\'e quantum dot arrays and concentrate on a detailed analysis of a structure with $N=9$ sites. We show that the structures reveal finite spin polarization for small twist angles and sufficiently strong interaction strength in the vicinity of half-filling and mainly above it. The origin of magnetization is two-fold. The main factor is direct exchange interaction due to the nonzero overlap of Wannier functions from neighboring moir\'e lattice sites. This leads to maximal spin polarization at half-filling that otherwise would not be expected on a triangular lattice. Additionally, there is a contribution to magnetization due to the mechanism proposed by Nagaoka as the effect of constructive interference between different paths of electrons moving in a spin-polarized background. This can lead to more stable magnetization away from half-filling (above it) in comparison to the half-filling case, which we observe in all considered structures. Nagaoka ferromagnetism is expected to vanish in a thermodynamic limit because the energy gap between the spin-polarized ground state and the excited states vanishes. In moir\'e quantum dot arrays this energy gap should still be finite, and, as we show,  is filling factor-dependent. Whether magnetization at half-filling or away from it is more stable depends on which of these two factors dominates. We have noticed that the shape of moir\'e quantum dot arrays determines the charge order for fillings away from charge neutrality, the magnetic properties are mainly determined by the twist angle. Appropriate choice of materials forming TMD heterostructure, size and shape of moir\'e quantum dot arrays, and the twist angle can allow designing magnetic nanostructures with a filling factor dependent magnetization.

\section*{ACKNOWLEDGMENTS}
The authors acknowledge helpful interactions with M. Zielinski.
PP acknowledges support from the Polish National Science Centre based on Decision No. 2021/41/B/ST3/03322. MK acknowledges support from the Polish National Science Centre based on Decision No. 2019/33/N/ST3/03137. Calculations were performed using the Wrocław Center for Networking and Supercomputing WCSS.

\bibliography{refs}

%\appendix
\onecolumngrid
\section*{Appendix}

\subsection*{Continuum model and Wannier functions from projection technique}
 The valley-projected continuum Hamiltonian for TMD moir\'e heterobilayers is written as \cite{FengchengHubbard}
\begin{align}
    H=-\frac{\hbar^2}{2m^*}{\bf k}^2+\Delta({\bf r}),
    \label{moir\'eHamiltonian}
\end{align}
where $m^* = 0.35 m_0$ \cite{FengchengHubbard,Nonlocal} is the effective mass of charge carriers in the valence band of the active layer and a modulation potential $\Delta({\bf r})=2V_{\rm m}\sum_{j=1,3,5}\cos({\bf b}_j\cdot{\bf r}+\phi)$ with ${\bf{b}}_j=4\pi/\sqrt{3}a_M\left(\cos\left(\pi j/3\right),\sin\left(\pi j/3\right)\right)$, belonging to the first shell of reciprocal lattice vectors with $V_{\rm m} = 25$ meV and $\phi = -94^{\circ}$, which determine the strength of the potential and the location of its minima, respectively. The Hamiltonian given by Eq. (\ref{moir\'eHamiltonian}) is diagonalized in a plane wave basis giving energies $E_n$ and eigenstates $|\Psi_n ({\bf k}) \rangle$. Bloch $|\Psi ({\bf k}) \rangle$ of the topmost valence band (we omit the band index) are
\begin{align}
    \ket{\Psi({\bf k})}=\sum_{\bf G}z_{{\bf k+G}}e^{i({\bf k}+{\bf G}){\bf r}},
\end{align}
where $z_{{\bf k+G}}$ are  expansion coefficients. 

We use a projection technique \cite{Cloizeaux1,Cloizeaux2,Vanderbilt} to obtain Wannier functions of holes localized on moir\'e superlattice sites.
We project trial wavefunction $|t_i ({\bf r})\rangle$, with corresponding Fourier transform $|t_i ({\bf k}) \rangle$ onto Bloch function $|\Psi ({\bf k}) \rangle$ of topmost valence band of moir\'e band structure. One has
 \begin{eqnarray}
|\gamma_i ({\bf k}) \rangle =  P({\bf k}) | t_i ({\bf k}) \rangle = |\Psi ({\bf k}) \rangle \langle \Psi ({\bf k}) | t_i ({\bf k}) \rangle   \nonumber
\end{eqnarray}
with  overlap 
 \begin{eqnarray}
S= \langle t_i ({\bf k})  | P({\bf k}) | t_i ({\bf k}) \rangle = \langle t_i ({\bf k})  | \Psi ({\bf k})   \rangle \langle \Psi ({\bf k}) | t_i ({\bf k}) \rangle  =  |\langle \Psi ({\bf k}) | t_i ({\bf k}) \rangle|^2   \nonumber
\end{eqnarray}
and orthogonal new quasi-Bloch states are
 \begin{eqnarray}
|\tilde{\Psi} ({\bf k}) \rangle = (S^{-1/2}) |\gamma ({\bf k}) \rangle = |\Psi ({\bf k}) \rangle \frac{\langle \Psi ({\bf k}) | t ({\bf k}) \rangle}{|\langle \Psi ({\bf k}) | t ({\bf k}) \rangle|} =  |\Psi ({\bf k}) \rangle e^{-i\theta_t({\bf k})},
\label{Psiktild}
\end{eqnarray}
which means that the projection technique is equivalent to appropriate $\theta_t({\bf k})$ phase fixing of Bloch states in order to have corresponding exponentially localized Wannier functions on lattice site $i$. We choose trial state $|t_i\rangle$ as a delta function (Gaussian-like trial wavefunction gives similar results) localized at a given site of a crystal lattice,
 \begin{eqnarray}
|t_i ({\bf r})\rangle = \delta ({\bf r} - {\bf \tau}_i )  \nonumber
\label{tau}
\end{eqnarray}
where ${\bf \tau}_i$ determines a position of a lattice site. Fourier transform of it is
 \begin{eqnarray}
|t_i ({\bf k}) \rangle = \frac{1}{\sqrt{N_G}} \sum_{{\bf G}} e^{i({\bf k}+{\bf G})({\bf r} - {\bf \tau}_i)}  \nonumber
\label{tauk}
\end{eqnarray}
where $N_G$ is a number of reciprocal basis vectors $\bf G$. One has an overlap with the Bloch state 
 \begin{eqnarray}
\langle \Psi ({\bf k}) | t ({\bf k}) \rangle =  \frac{1}{\sqrt{N_G}} \sum_{{\bf G},{\bf G}'}  
z^*_{{\bf k+G}}  \int d{\bf r} e^{-i({\bf k}+{\bf G}'){\bf r}}  e^{i({\bf k}+{\bf G})({\bf r} - {\bf \tau}_i)}  
= \frac{V_{cell}}{\sqrt{N_G}} \sum_{{\bf G}}  
z^*_{{\bf k+G}}  e^{-i({\bf k}+{\bf G}){\bf \tau}_i}, \nonumber    
\end{eqnarray}
where $V_{cell}$ is the unit cell volume. After the above procedure of fixing the phase of the topmost valence band, eigenstates corresponding localized Wannier functions can be obtained
\begin{align}
     %\Psi_{{\bf R}}({\bf r})
     \ket{{\bf R}}=\frac{1}{\sqrt{N}}\sum_{\bf k}e^{-i\,{\bf k}\cdot{\bf R}}\ket{\tilde{\Psi} ({\bf k})},
     %\sum_{\bf G}z^n_{\bf k+G}e^{i\,({\bf k+G})\cdot{\bf r}},
\end{align}
where ${\bf R}$ are Wannier center positions on a moir\'e triangular lattice and $N$ is the number of moir\'e unit cells. Four-center real-space Coulomb matrix elements are 
\begin{align}
\label{Coulomb}
    \braket{{\bf R}_i,{\bf R}_j|V|{\bf R}_k,{\bf R}_l}=\frac{1}{N^2}\sum_{\substack{{\bf k}_i,{\bf k}_j\\{\bf k}_k,{\bf k}_l}}e^{i ({\bf k}_i\cdot{\bf R}_i+{\bf k}_j\cdot{\bf R}_j-{\bf k}_k\cdot{\bf R}_k-{\bf k}_l\cdot{\bf R}_l)} \braket{\tilde{\Psi} ({\bf k}_i),\tilde{\Psi} ({\bf k}_j)|V|\tilde{\Psi} ({\bf k}_k),\tilde{\Psi} ({\bf k}_l)},
\end{align}
with $V=\frac{e^2}{4\pi\epsilon_0 \epsilon|{\bf r}_1-{\bf r}_2|} $, with $e$ as electric charge and $\epsilon_0$ is the vacuum permittivity,  ${\bf r}_i$ is position of $i$-th particle, and two-center integrals are: $U_0 $ ($i=j=k=l$), $U_1$ ($i=l$, $j=k$ and $i,j$ are nearest neighbors), $X_1 $ ($i=k$, $j=l$ and $i,j$ are nearest neighbors), $A$ ($i=j=k$, and $i,l$ are nearest neighbors). We also take $U_n = U_1/r_n$, where $r_n$ is a distance to $n$-th nearest neighbors, assuming $r_1=1$ for nearest neighbors. We perform calculations of Coulomb elements in reciprocal space using the Bloch state defined by Eq. \ref{Psiktild}.  Tight-binding hopping integrals are given by  \begin{eqnarray}
t_{n} =\frac{1}{N}\sum_{\bf k}  e^{-i{\bf k}({\bf R}_i - {\bf R}_j)}E_{{\bf k}},
\label{EnR}
\end{eqnarray}
where $n=1$ for $(i,j)$ nearest neighbors and $n=2$ for $(i,j)$ next nearest neighbors, and $n=3$ for $(i,j)$ next-next nearest neighbors. To confirm the validity of our real-space parameters we have compared exact diagonalization results starting from a continuum model and with real space parameters obtaining satisfactory agreement for a periodic system with $N=9$ unit cells ($3\times 3$ momentum space mesh) and $N_{el}=9$ particles. Generalized Hubbard model parameters are presented in table \ref{table:1}.

\begin{table}[h]
\centering
\begin{tabular}{||c | c c c c c c c||} 
 \hline
 $\theta$ & $t_1$ & $t_2$ & $t_3$ & $U_0$ & $U_1$ & $X_1$ & $A_1$ \\ [0.5ex] 
 \hline\hline
 2.5 & 0.653 & -0.024 & -0.0147 &1086.925 & 158.632 & 0.308 & -3.301 \\ 
 3.5 & 3.341 & -0.389 & -0.217 &1143.612 & 229.822 & 4.057 & -4.135 \\
 5.0 & 10.351 & -2.066 & -1.032 &1232.254 & 336.887 & 17.387 & 10.006 \\ [1ex] 
 \hline
\end{tabular}
\caption{Generalized Hubbard model parameters for different twist angles $\theta$. All parameters are in meV and interaction parameters for $\epsilon=1$.}
\label{table:1}
\end{table}

\end{document}